\documentclass[journal]{IEEEtran}

\usepackage{graphicx,amssymb,lineno}
\usepackage{amsmath,amsfonts,amssymb}

\usepackage{algorithmic}
\usepackage[usenames]{color}
\usepackage{float}

\usepackage[linesnumbered,ruled,vlined]{algorithm2e}

\newcommand*{\TitleFont}{%
	\usefont{\encodingdefault}{\rmdefault}{}{n}%
	\fontsize{20}{20}%
	\selectfont}

\usepackage{graphicx,graphics,color,epsfig,subfigure,graphpap,rotate}
\usepackage{times, verbatim, subfigure, epsfig, graphicx, latexsym, amsmath}
\usepackage{url}
\usepackage{subfigure}
\begin{document}
\title{\TitleFont Robust Transmission Scheduling for UAV-assisted Millimeter-Wave Train-Ground Communication System}

\author{Yunhan~Ma,
	    Yong~Niu,~\IEEEmembership{Member,~IEEE},
	    Zhu~Han,~\IEEEmembership{Fellow,~IEEE},
	    Bo~Ai,~\IEEEmembership{Fellow,~IEEE},
	    Kai~Li,~\IEEEmembership{Member,~IEEE},
	    Zhangdui~Zhong,~\IEEEmembership{Fellow,~IEEE},
		and Ning~Wang,~\IEEEmembership{Member,~IEEE}

\thanks{Copyright (c) 2015 IEEE. Personal use of this material is permitted. However, permission to use this material for any other purposes must be obtained from the IEEE by sending a request to pubs-permissions@ieee.org. This study was supported by the National Key Research and Development Program under Grant 2021YFB2900301; in part by National Key R\&D Program of China (2020YFB1806903); in part by the National Natural Science Foundation of China Grants 61801016, 61725101, 61961130391, and U1834210; in part by the State Key Laboratory of Rail Traffic Control and Safety (Contract No. RCS2021ZT009), Beijing Jiaotong University; and supported by the open research fund of National Mobile Communications Research Laboratory, Southeast University (No. 2021D09); in part by the Fundamental Research Funds for the Central Universities, China, under grant number 2022JBQY004 and 2022JBXT001; and supported by Frontiers Science Center for Smart High-speed Railway System; in part by the Fundamental Research Funds for the Central Universities 2020JBM089; in part by the Project of China Shenhua under Grant (GJNY-20-01-1); and in part by NSF CNS-2107216 and CNS-2128368.}

 \thanks{Y. Ma, B. Ai, Z. Zhong are with the State Key Laboratory of Rail Traffic Control and Safety, Beijing Jiaotong University, Beijing 100044, China, and also with the Beijing Engineering Research Center of High-speed Railway Broadband Mobile Communications, Beijing Jiaotong University, Beijing 100044, China (e-mails: 21120110@bjtu.edu.cn, boai@bjtu.edu.cn, zhdzhong@bjtu.edu.cn).} 
 
 \thanks{Y. Niu is with the State Key Laboratory of Rail Traffic Control and Safety, Beijing Jiaotong University, Beijing 100044, China, and also with the National Mobile Communications Research Laboratory, Southeast University, Nanjing 211189, China (e-mail: niuy11@163.com).}
 
 \thanks{Z. Han is with the Department of Electrical and Computer Engineering at the University of Houston, Houston, TX 77004 USA, and also with the Department of Computer Science and Engineering, Kyung Hee University, Seoul, South Korea, 446-701 (e-mail: hanzhu22@gmail.com).}
 
 \thanks{K. Li is with the Real-Time and Embedded Computing Systems Research Centre (CISTER), Porto, Portugal.}

 \thanks{N. Wang is with the School of Information Engineering, Zhengzhou University, Zhengzhou 450001, China (e-mail: ienwang@zzu.edu.cn).}
 
}
\maketitle
	
\begin{abstract}
With the explosive growth of mobile data, the demand of high-speed railway (HSR) passengers for broadband wireless access services urgently needs the support of ultra-high-speed scenario broadband wireless communication. Millimeter-wave (mmWave) can achieve high data transmission rates, but it is accompanied by high propagation loss and vulnerability to blockage. To address this issue, developments of directional antennas and unmanned aerial vehicles (UAVs) enhance the robustness of the mmWave train-ground communication system. In this paper, we propose a UAV and MRs relay assistance (UMRA) algorithm to effectively overcome link blockage, which can maximize the number of transmission flows on the premise of meeting QoS requirements and channel qualities. First, we formulate a mixed integer nonlinear programming (MINLP) problem for UAV trajectory design and transmission scheduling in the full-duplex (FD) mode. Then, in UMRA, the relay decision algorithm and transmission scheduling algorithm based on graph theory are proposed, which make a good tradeoff between computation complexity and system performance. Extensive simulation results show that a suitable UAV position will greatly improve the performance of the UMRA algorithm and make it close to the optimal solution. Compared with the other two existing benchmark schemes, with the high channel quality requirements and large-area blockage, UMRA can greatly improve the number of completed flows and system throughput.


\end{abstract}
\section{Introduction}\label{S1}
Nowadays, high-speed railway is developing rapidly. By 2019, the global high-speed railway can provide travel services for 1.7 billion people every year \cite{development}. So far, the speed of Japan’s latest bullet train ``Falcon'' and German ice train exceeds 300 kilometers per hour \cite{japan}. China has built high-speed rail lines with a speed of more than 350 kilometers per hour \cite{china}. With the constantly refresh of high-speed railway speed, passengers hope to obtain the support of broadband wireless communication in the ultra-high-speed scene. The demand data rate of each carriage will reach 0.5 – 5 Gbps in the future \cite{speed}. In order to realize the needs of users, such as instantaneous music and real-time online streaming media, we need to study the high-speed train-ground communication.
	
At present, wireless data traffic is growing at a rate of more than 50\% per user every year \cite{yewu}. With the problem of spectrum shortage, millimeter-band (mm-band) can provide multigigabit communication services for mobile user equipments (UEs), support bandwidth-intensive multimedia applications, and become an advanced technology of the 5G mobile network \cite{deve},  \cite{usa}. Its characteristics of narrow beam and rich bandwidth resources can also be applied to train-ground communication systems.

Due to the high frequency and short wavelength of millimeter-wave (mmWave), the diffraction ability of electromagnetic wave is very weak for obstacles whose size is significantly larger than the wavelength. The higher the frequency, the more sensitive the link is to obstacles.	
At the same time, mmWave causes a lot of loss when propagating at high frequencies. According to the Friis free space formula, the free space loss is directly proportional to the square of carrier frequency, and the loss of 60-GHz signal is 28 dB higher than that of 2.4-GHz \cite{loss}. In the process of propagation, rain attenuation and atmospheric molecular absorption limit the distance of mmWave communication. A large number of antennas are placed in the transceiver, and the beamforming techniques are used to obtain higher gain and extend the communication range. 
	
The antenna arrays are active. By controlling the signal phase transmitted by the antenna elements, the beam can be converted to any direction. If the beam pointing angle is aligned, high gain can be obtained in this direction. At the same time, large scale antenna array can resist path loss in mmWave band \cite{ann} and improve mmWave spectral efficiency and propagation range \cite{NR}. 
We consider the combination of full-duplex (FD) communication and antenna array. Under the given bandwidth, we can double the throughput while achieving high gain \cite{FDHD}. However, the cost of improving spectral efficiency of FD communication is to overcome the self-interference (SI) from the transmitter \cite{daijia}. The transmitting antenna and receiving antenna need high-quality decoupling \cite{jieou}. 
	
Stronger link losses require the support from new relays to create better channel conditions, such as unmanned aerial vehicle (UAV). The UAV can hover above the BS service area in a quasi-static state. Because it is not affected by many scatterers, the air-to-ground is mostly the LOS channel. The introduction of UAV can improve NLOS link to LOS link, reduce link loss and realize highly directional transmission, which is convenient to maintain a lasting communication state in HSR scenario \cite{uav}. At the same time, UAV has the characteristics of flexibility, rapid deployment on demand and low cost. It can quickly establish communication links and flexibly adjust the location of UAV \cite{uavfeacture}, which can overcome link blockage in HSR scenario and improve communication coverage. However, for safety and energy consumption, the weight of UAV used for communication is usually no more than 25 kg. Combined with the characteristics of mmWave ultrashort wavelength, the micro-sized antennas can be better packaged on UAVs with limited loads \cite{fengzhuang}. UAV-assisted mmWave communication can use the 60-GHz unlicensed ISM frequency band, which can reach a peak rate of 10 Gbit/s  \cite{sulv}. At the same time, the directional beam and narrow beam width of mmWave can effectively resist interference and eavesdropping, making the information transmission of train-ground communication system more secure \cite{ganrao} \cite{qieting} \cite{anquan} \cite{miyao}.	


In the HSR scenario, a large number of passenger data demands can not be met on trains with large passenger flow, at the time of train arrival and peak hours. We can introduce mmWave into HSR \cite{mmwavehsr1} \cite{mmwavehsr2} \cite{mmwavehsr3} \cite{mmwavehsr4} to solve the problem of insufficient spectrum. With the help of UAV, using its air-to-ground channel characteristics \cite{uav} \cite{airground} and the advantage of flexible adjustment through track design, we can quickly establish highly directional LOS links with each communication node to ensure channel quality and solve the problem of link blockage during mmWave transmission. Although the coverage of mmWave is small, it is very useful to improve the capacity of some hot spots and is suitable for areas with large traffic demand. The UAV in this paper hovers at a certain position in the air and does not follow the high-speed rail. It can play a role of relay assistance in hot spots and time periods. With the high-speed transmission characteristics of mmWave, the UAV can complete the data flow transmission of high-speed trains in a short time under its own coverage. The introduction of the FD communication mode, combined with the mmWave technology, greatly improves the communication rate and communication capacity. The contributions of this paper can be summarized as follows.
\begin{itemize}
\item In order to reasonably allocate relays in links are blocked and do not meet the transmission requirements, a low complexity UAV and MRs relay assistance (UMRA) heuristic algorithm based on graph theory is proposed. We use UAV and MRs installed on the rooftop of the train to replan the link in the case of large-area link blockage. On the premise of meeting the quality of service (QoS) requirements and channel qualities, the system throughput and the number of completed flows are greatly improved.
	
\item We combine the ultra-high-speed scene with UAV and mmWave technology. FD technology is exploited to double the network capacity and carrying capacity. By optimizing UAV's deployment, we can enhance the communication coverage and significantly improve the efficiency of mmWave train-ground communication.

\item Comparing the results of the heuristic UMRA algorithm with the exhaustive search, it is found that the results obtained by the UMRA algorithm are approximate to the optimal solution. Extensive simulation results show that its performance is much higher than the other two benchmark schemes.
\end{itemize}

The rest of this paper is arranged as follows. The related work is discussed in Section~\ref{S2}. In Section~\ref{S3}, the system model is established to analyze the blockage situation, and the UAV-assisted relay mode is proposed. In Section~\ref{S4}, the mathematical model is established for blocked links and transmission scheduling. In Section~\ref{S5}, the heuristic UMRA algorithm based on graph theory is proposed. In Section~\ref{S6}, the algorithm is compared with benchmark schemes and the optimal solution. Finally, we conclude this paper in Section~\ref{S7}.
\section{Related Work}\label{S2}
Millimeter-wave has been widely used in high-speed railway systems, such as Shinkansen in Japan, maglev train in Shanghai and so on \cite{world}. The rapid development of high-speed railway urgently needs the 5G-related technical support. At present, 5G mainly studies 28 GHz, 38 GHz (with 3-4 GHz available spectrum) and E-band (71-76 GHz and 81-86 GHz) (with 10 GHz available spectrum) \cite{5G}. Due to the great potential of mmWave, ECMA, IEEE 802.15.3 task group 3C, IEEE 802.11ad, WiGig and other organizations have standardized \cite{mmwave}. The transmission frequency of CMOS transistors can reach hundreds of gigahertz, which is much higher than the carrier frequency of 60 GHz. The progress of CMOS RF integrated circuit technology can achieve higher gain and output power, so that mmWave can provide unprecedented data rate \cite{CMOS}.
	
Millimeter-wave technology is relatively mature and can be introduced into the field of ultra-high-speed.  Facing the new challenge of supporting high mobility, some literatures have studied the channel characteristics of typical high-speed rail scenarios in 5G mm-band. They analyzed channel parameters for urban, rural and tunnel environments and guided the design of typical mmWave communication systems \cite{mmhsr}. However, Doppler shift is a critical issue in applying mmWave to HSR. Chen $ et\ al. $ \cite{Zshape} analyzed the influence of Doppler on the channel characteristics in the HSR scenario. The change of Doppler frequency is large near the BS and will sweep from the maximum to the minimum frequency quickly. In order to avoid related problem, Gong $ et\ al. $ \cite{aided}  designed a data-aided Doppler estimation and compensation algorithm using the channel matrix model of second-order Taylor expansion. Xiong $ et\ al. $ \cite{LSTM}  proposed a Doppler prediction algorithm based on long short-term memory (LSTM) neural network, which can learn the regularity of Doppler shift through pretraining and tracking training. Song $ et\ al. $ \cite{scfde}  designed a SC-FDE frame structure to estimate the problems of Doppler effect and enhance the feedback delay sensitivity in high-speed railways, which can achieve high system throughput. Song $ et\ al. $ \cite{mmwavehsr1} applied mmWave to HSR system to solve the poor propagation characteristics of mmWave and the particularity of high-speed railway scene.
	
To solve the problem of mmwave blocked links, many literatures have proposed the use of mobile access points (APs) for relaying. Some literatures used multi-hop relay to forward services of interrupted links to solve the problem of blockage sensitivity \cite{relaymac}, \cite{LI}. Chen $ et\ al. $ \cite{tworelay} proposed a two-hop relay scheme to improve the performance of WLAN uplink channel access under the IEEE 802.11ad standard. The scheme used two-hop high-quality links to relay direct communication links with poor channel qualities, which significantly improved the system throughput. Pan $ et\ al. $ \cite{ap} proposed an enhanced handover scheme using mobile relays. The user equipment has connected to the onboard mobile relay, and the mobile relay established a backhaul link with the service host eNB to support the communication service in the LTE-A high-speed railway system.
	
3GPP \cite{uavchannel} has carried out experiments on the communication channel of the UAV system. The results showed that LOS channel dominated air to ground channel in many practical scenarios, especially in rural areas or medium height. Nowadays, UAV cooperative control is widely used in communication networks. Wu $ et\ al.$ \cite{yingji} established a network model using UAV networking technology and confirmed that UAVs can be applied to high-speed railway emergency communications. 
Ruan $ et\ al.$ \cite{muecd} proposed a multi-UAV coverage model for energy-saving communication based on game theory, which used the muecd-sap algorithm to perform motion and power control, so as to ensure the suboptimal energy efficiency coverage deployment.

At present, UAV is widely used in mmWave wireless communication. Gao $ et\ al. $ \cite{UM1} considered a mission driven multi-UAVs network with mmWave transmission in the literature, and proposed an optimization scheme of NOMA-grouping-aware fast transmit beamforming based on deep learning. It can improve the coverage of mmWave-NOMA transmissions in highly dynamic multi-UAVs networks. Li $ et\ al. $ \cite{UM2} proposed a joint Doppler shift and channel estimation method for the mmWave communication system of an UAV equipped with a large ULA array. Chang $ et\ al. $ \cite{UM3} proposed a new integrated scheduling method of sensing, communication and control for mmWave communication in UAV networks to realize data transmission of the backhaul from UAV to the ground BS.
	
You $ et\ al.$ \cite{fdhd} analyzed and compared the traversal capacity and interruption performance of FD and half-duplex (HD) relay transmission in high-speed railway. When the carrier penetration loss CPL exceeded 5 dB, the FD scheme is superior to the HD scheme in terms of system reliability and efficiency. Therefore, we intend to introduce FD communication into HSR. Based on the competitive graph, Ding $ et\ al.$ \cite{fdmm} combined FD communication with mmWave and proposed a QoS-aware FD concurrent scheduling algorithm to ensure the high-speed transmission of flows. Wang $ et\ al.$ \cite{uavfd} combined UAV with FD technology to design an efficient spectrum sharing method for D2D communication between aerial UAV and ground to maximize system throughput at fixed transmitting power. Zhu $ et\ al.$ \cite{fduavmm} applied FD and UAV to the mmWave communication network to jointly optimize UAV position, beamforming and power control to maximize achievable rates. 
		\begin{figure}[t]
		\begin{minipage}[t]{1\linewidth}
			\centering
			\includegraphics[width=1\columnwidth]{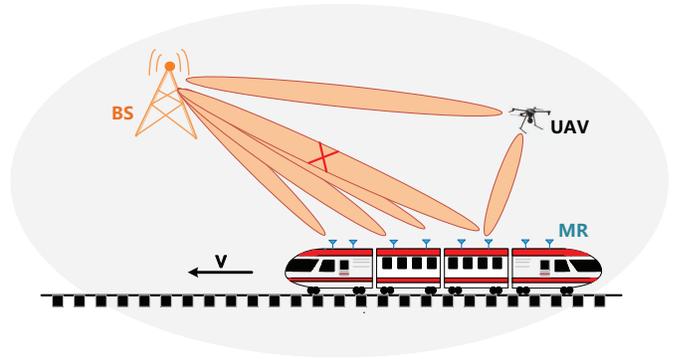}
		\end{minipage}%
		\caption{Millimeter-wave train-ground communication system model}
		\label{fig:hsr} 
		\vspace*{-3mm}
	\end{figure}
	
With flexible networking and communication reliability, UAV is widely used in disaster areas, hot spots and complex terrains \cite{Uav}, providing new degrees of freedom for the wireless communication system. However, most studies are aimed at transmission scheduling in WLAN or cellular scenarios. UAVs are rarely used in mobile scenarios such as in-vehicle communications and mobile networks. There is a lack of research on transmission scheduling in the high-speed train operation and FD robust transmission mechanism based on mmWave train-ground communication in ultra-high-speed scenarios. Therefore, this paper applies UAVs to the high-mobility mmWave train-ground networks to assist transmission in mobile environments. As an air relay, the UAV realizes the air access of the BSs to mobile users.
\section{System Model}\label{S3}

We consider a scenario of mmWave train-ground communication system, as shown in Fig. \ref{fig:hsr}. There are $ F $ MRs in fixed positions, which are evenly distributed on the top of the train to realize the communication between the train and the BS. BS is fixed outside the railway track and a UAV hovers in the air. The UAV is located within the coverage of the BS on the orbital side for robust relay transmission. All equipments adopt the FD communication mode which only consider downlink transmission. The flow here refers to the data flow with traffic to be transmitted, which is sent to the MRs on the top of the train by the BS. In the most complex case, the number of flows is equal to the number of MRs, that is, each MR has its traffic to be received from BS. All links involved in Fig. \ref{fig:hsr} adopt mmWave frequency transmission with  $ f  = \rm{28\ GHz} $ , in which BS, MRs and UAV are equipped with steerable directional antennas, allowing them to aim at relevant users to obtain higher antenna gain. The high-speed movement of the train and the high carrier frequency of mmWave will result in serious Doppler shift. It is more likely to cause packet loss, long delay, inter symbol interference, signal quality degradation at the receiver, etc. The Doppler shift can be expressed as
\begin{equation}
	{{f}_{D}}=\frac{v}{\lambda }\cos \theta =\frac{v}{c}{{f}_{c}}\cos \theta,
\end{equation}
where $ f_{c} $	is the carrier frequency, $ \lambda $ is the corresponding wavelength, $ v $ and $ c $ are the velocity of train and the speed of light respectively. $ \theta $ is the arrival direction angle of the LOS path to MR. When $ \theta $ is 0, the maximum Doppler shift can be obtained as
\begin{equation}
	{{f}_{D,\max }}=\frac{v}{c}{{f}_{c}}.
\end{equation}
Therefore, the faster the speed is, the greater the frequency shift is. $ {{f}_{D,\max }} $ is about 7800Hz at the speed of 300km/h. In the high-speed rail scenario, the BS is closer to the train, the angle between radio propagation and the direction of train movement is very small, which will lead to high Doppler frequency shift.  The shape of Doppler spectrum is narrow. When the train passes through the BS, the Doppler shift of the LOS path sweeps from positive to negative, which is characterized by a ``Z'' shape, rapidly sweeping from the maximum to the minimum \cite{Zshape}. During the 2km measurement process, the high-speed railway running at a constant speed finally corresponds to an almost constant Doppler shift value \cite{2km}. This is because the train runs along a constant, regular and predictable route at a relatively stable speed, the Doppler shift experienced by the train at any position is also predictable. Given the powerful processing ability of MR, we deploy a neural network on one of the MRs, which uses a machine learning method based on the reference signal received power (RSRP) values to predict Doppler shift \cite{buchang}. RSRP is measured every certain distance. After standardization and noise preprocessing, these RSRP data sets are used as the basic data of machine learning. Through training, the neural network becomes the function \textbf{G(·)} of the relative Doppler shift estimator (RDSE). When $ s\left( d \right) $ is input, the relative Doppler shift $ {{f}_{D,rel}}\left( d \right) $ can be estimated as
\begin{equation}
	{{f}_{D,rel}}\left( d \right)=G\left( s\left( d \right) \right).
\end{equation}
Where $ s\left( d \right) $ is expressed as
\begin{equation}
	s\left( d \right)=\left[ x\left( d-\frac{l}{2} \right),\cdots ,x\left( d \right),x\left( d-\frac{l}{n} \right),\cdots ,x\left( d+\frac{l}{2} \right) \right],
\end{equation}
where $ l $ is the measurement range, and select $ n+1 $ equidistant  RSRP measurements near node $ d $ to generate the input set $ s\left( d \right) $. This is because different locations may correspond to the same RSRP value. If a single RSRP value is input into the neural network, different locations will produce the same Doppler shift value, resulting in a one-to-many mapping problem. Therefore, we use multiple surrounding RSRP values as input to create a unique RSRP input set for each different HSR location. The relative shift value $ {{f}_{D,rel}}\left( d \right) $ is between -1 and 1. Combined with the maximum offset, the Doppler shift value $ {{f}_{D}}\left( d \right) $ can be obtained as 
\begin{equation}
	{{f}_{D}}\left( d \right)={{f}_{D,rel}}\left( d \right)*{{f}_{D,\max }}={{f}_{D,rel}}\left( d \right)*\frac{v}{c}{{f}_{c}}.
\end{equation}

\begin{figure}[t]
		\begin{minipage}[t]{1\linewidth}
			\centering
			\includegraphics[width=0.9\columnwidth]{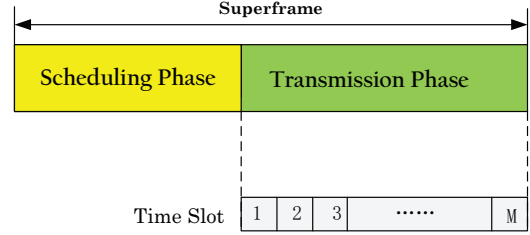}
		\end{minipage}%
		\caption{MAC transmission frame structure}
		\label{fig:mac} 
		\vspace*{-3mm}
	\end{figure}

The system adopts MAC transmission frame structure based on time division multiple access (TDMA) \cite{MAC}. Time is divided into a series of nonoverlapping superframes, each superframe is composed of scheduling stage and transmission stage, as shown in Fig. \ref{fig:mac}. When the train-ground communication link is interrupted, the repeater will notify the BS of the interruption information in the scheduling stage. According to the interruption information, the BS completes the relay transmission via UAV and MRs. All links that need to be relayed are also scheduled by the BS. Finally, all links are scheduled in the transmission stage.
	
\begin{figure}[t]
	\begin{minipage}[t]{1\linewidth}
		\centering
		\includegraphics[width=0.7\columnwidth]{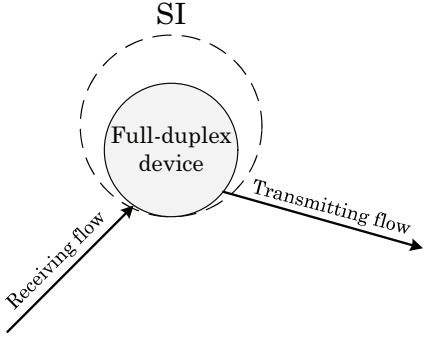}
	\end{minipage}%
	\caption{Self-interference model of full-duplex communication}
	\label{fig:SI} 
	\vspace*{-3mm}
\end{figure}

All nodes working in FD communication mode may exist self-interference (SI), which means that the signal sent by the transmitter is received by the receiver of the same node \cite{SI}, as shown in Fig. \ref{fig:SI}. When the transmitting power is ${{P}_{t}}$, the SI model can be formulated as
	\begin{equation}
		{{I}_{s}}=\beta {{P}_{t}},
	\end{equation}
where $ \beta $ represents the level of SI cancellation. After the SI is eliminated, the two-hop links using relay can transmit at the same time without being affected by each other. 
	
Suppose that there are $ F $ flows to be transmitted in the system. There are four channels in this paper, including the channel from BS to MR, the channel from BS to UAV, the channel from UAV to MR and the channel from MR to the MR. Due to the great link loss of NLOS channel, data loss and reception delay will be caused, we intend to use relay to turn it into LOS channel, which can meet the requirements of millimeter wave transmission link.All links in this paper adopt the mmWave LOS path loss model \cite{pathloss}. Therefore, For each flow of $f$, its received power ${{P}_{r}}\left( f \right)$ can be expressed as
\begin{equation}
		{{P}_{r}}\left( f \right)={{k}_{0}}{{P}_{t}}{{G}_{t}}\left( {{s}_{f}},{{t}_{f}} \right){{G}_{r}}\left( {{s}_{f}},{{t}_{f}} \right)d_{{{s}_{f}}{{t}_{f}}}^{-\alpha },
\end{equation}
where ${k}_{0}$ is a constant proportional to ${{\left( \frac{\lambda }{4\pi } \right)}^{2}}$, $\lambda$ is the wavelength of transmission signal. ${s}_{f}$ and ${t}_{f}$ represent the transmitter and receiver of flow $f$. ${{G}_{t}}\left( {{s}_{f}},{{t}_{f}} \right)$ and ${{G}_{t}}\left( {{s}_{f}},{{t}_{f}} \right)$ are the antenna gain of the transmitter and receiver. The distance between ${s}_{f}$ and ${t}_{f}$ is $d_{{{s}_{f}}{{t}_{f}}}$, $\alpha$ is the path loss index.
	
For the receiver of flow $f$, its signal-to-interference-plus noise ratio (SINR) can be expressed as
	\begin{equation}
		\Gamma \left( f \right)=SIN{{R}_{f}}=\frac{{{P}_{r}}\left( f \right)}{{{N}_{0}}W+\gamma {{I}_{s}}\left( f \right)},
	\end{equation}
where ${{N}_{0}}$ is the one-sided noise power spectral density of the Gaussian channel. $ W $ is the channel bandwidth, and $\gamma$ is a sign whether SI exists under FD communication. When the flow selects relay mode for transmission, it needs to be transmitted twice to reach the destination. The first hop is from the BS to the relay node, and the second hop is from the relay node to the destination node. Because all devices (MRs and UAV) transmit in FD mode, they can transmit and receive at the same time. While being received the first hop, the relay node is also sending traffic to the destination node. For a device transmitting and received at the same time, the transmitting power of the second hop has SI with the received power of the first hop. The SI affects the received power and transmitting rate of the first hop. Although the receiver of the second hop works in FD mode, there is no transmitting link, so the receiver of the second hop is not affected by SI. Therefore, SI can exist only in the first hop of a link using relay, at this time $\gamma=1$, it is 0 in other cases. ${{I}_{s}}\left( f \right)$ refers to the FD interference generated by its own transmitter to the receiver.
Then the receiving rate of link $f$ can be calculated according to the Shannon's channel capacity as
\begin{equation}
		{{R}_{f}}=\eta W{{\log }_{2}}\left[ 1+\Gamma \left( f \right) \right]=\eta W{{\log }_{2}}\left[ 1+\frac{{{P}_{r}}\left( f \right)}{{{N}_{0}}W+\gamma \beta {{P}_{t}}\left( f \right)} \right],
\end{equation}
where $\eta \in \left( 0,1 \right)$ is the efficiency of transceiver design. The f-th MR node in the system is represented by $ MR_{f} $, and $ MR_{f-1} $ represents the node on the left side of $ MR_{f} $. We can calculate the transmission rate from the BS to $M{{R}_{f}}$ as
\begin{equation}
	R_{BS,M{{R}_{f}}}^{s}=\eta W{{\log }_{2}}\left[ 1+\frac{{{P}_{r}}\left( M{{R}_{f}} \right)}{{{N}_{0}}W} \right].
\end{equation}

Next, the transmission rate of any flow using relay is calculated. It is assumed that the blocked flow $ f $ is relayed using the node $ M{{R}_{f-1}} $ on the left.
The equipment utilizes FD communication, and has self-interference. For the first hop of the link $ f $ using relay, the receiving power of the relay node $M{{R}_{f-1}}$ is affected by the FD self-interference at the transmitter. Therefore, $\gamma =1$; The second hop has no FD interference, so that $\gamma =0$. SINRs and the rates of two hops can be calculated, respectively. For the first hop, we can calculate as
\begin{equation}
	\begin{aligned}
\Gamma _{BS,M{{R}_{f}}}^{{{l}_{1}}}&=\frac{{{P}_{r}}\left( M{{R}_{f-1}} \right)}{{{N}_{0}}W+\beta {{P}_{t}}\left( M{{R}_{f-1}} \right)}.\\
	R_{BS,M{{R}_{f}}}^{{{l}_{1}}}&=\eta W{{\log }_{2}}\left[ 1+\frac{{{P}_{r}}\left( M{{R}_{f-1}} \right)}{{{N}_{0}}W+\beta {{P}_{t}}\left( M{{R}_{f-1}} \right)} \right].
    \end{aligned}
\end{equation}
SINR and the transmission rate of the second hop are
\begin{equation}
		\begin{aligned}
	\Gamma _{BS,M{{R}_{f}}}^{{{l}_{2}}}&=\frac{{{P}_{r}}\left( M{{R}_{f}} \right)}{{{N}_{0}}W }.\\
		R_{BS,M{{R}_{f}}}^{{{l}_{2}}}&=\eta W{{\log }_{2}}\left[ 1+\frac{{{P}_{r}}\left( M{{R}_{f}} \right)}{{{N}_{0}}W }\right].
		\end{aligned}
\end{equation}
The final SINR and transmission rate of the link depend on a hop with worse channel conditions. Therefore, SINR and the transmission rate of the link using the left-end MR for relay are
\begin{equation}
    	\begin{aligned}
	\Gamma _{BS,M{{R}_{f}}}^{l}&=\min \left\{ \Gamma _{BS,M{{R}_{f}}}^{{{l}_{1}}},\Gamma _{BS,M{{R}_{f}}}^{{{l}_{2}}} \right\}.\\
		R_{BS,M{{R}_{f}}}^{l}&=\min \left\{ R_{BS,M{{R}_{f}}}^{{{l}_{1}}},R_{BS,M{{R}_{f}}}^{{{l}_{2}}} \right\}.
		\end{aligned}
\end{equation}

Due to the weak diffraction ability of mmWave, it is easy to be blocked by obstacles. We try to incorporate the UAV into this communication system. If some MRs in the communication system are blocked, UAV or its left and right adjacent MRs should be considered for relaying and indirect forwarding to the blocked nodes. Then, each node receives data flows from the BS, nearby MRs and UAV with a certain probability. When several consecutive MRs are blocked, the adjacent MRs relay cannot be used. At this time, UAV assistance plays an important role in effectively enhancing communication robustness. In addition, this paper proposes a method based on graph theory for blocked links, which is convenient to quickly select the appropriate relay mode, so as to maximize the number of flows that meet QoS requirements and channel qualities within the specified number of time slots.

\section{Problem Formulation}\label{S4}
It is assumed that there are $ F $ MR nodes and $ F $ transmission links corresponding to them. All devices exploit the FD communication mode. Considering downlink transmission, each flow can only use one specific way to complete the transmission. To maximize spatial reuse, the optimal scheduling should complete the transmission of most data flows within the specified number of time slots. We utilize the MAC frame structure with $ M $ transmission slots, make a schedule that represents a superframe. It is assumed that it has $ k $ stages, each stage contains several continuous CTAs and can only serve one MR. According to the distance from the BS to the MR, the received power of MR using the direct link can be obtained as
	\begin{equation}
		{{P}_{r}}\left( f \right)={{k}_{0}}{{P}_{t}}{{G}_{t}}\left(BS,f\right){{G}_{r}}\left(BS,f\right)d_{BS,f}^{-\alpha }.
	\end{equation}

		\begin{figure*}[t]
	\centering
	\includegraphics[width=17cm,height=5.5cm]{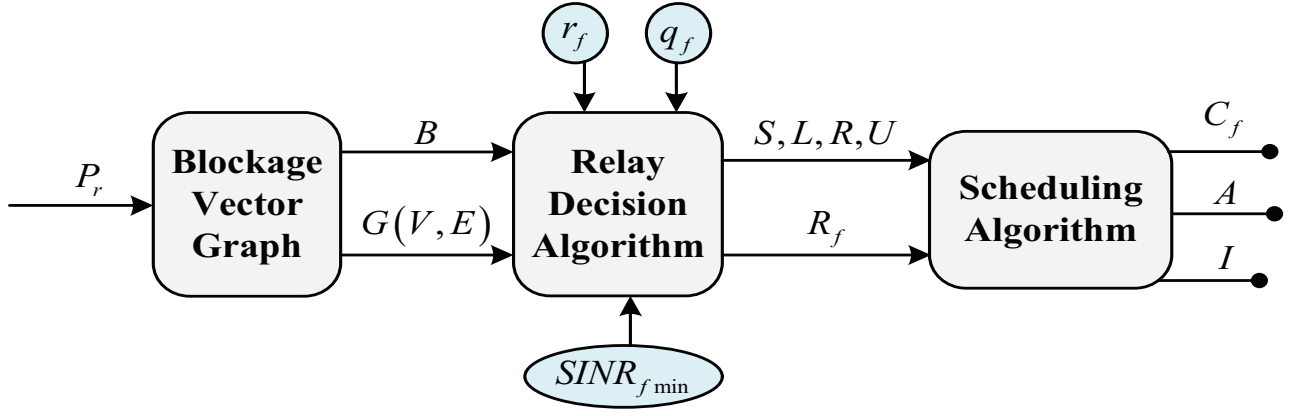}
	\caption{Flow chart of the heuristic UMRA algorithm}
	\label{fig:liucheng} 
	\vspace*{-3mm}
\end{figure*}	
Each flow has its own QoS requirement ${{q}_{f}}$, the minimum transmission rate required for the actual transmission link is 
	\begin{equation}
		{{r}_{f}}=\frac{{{q}_{f}}\left( {{T}_{s}}+M*\Delta t \right)}{M*\Delta t}.
	\end{equation}
If the actual transmission rate of flow $ f $ is ${{R}_{f}}$, the actual number of time slots to be occupied in the transmission process is
	\begin{equation}
		{{\delta }_{f}}=\left\lceil \frac{{{q}_{f}}\left( {{T}_{s}}+M*\Delta t \right)}{{{R}_{f}}*\Delta t} \right\rceil.
	\end{equation}

Next, define four binary variables for each flow $f$: $a_{f}^{s}$, $a_{f}^{l}$, $a_{f}^{r}$ and $a_{f}^{u}$. They indicate whether the flow is transmitted by direct link, left-end relay link, right-end relay link and UAV relay link, respectively. If this mode is used, the value is 1, otherwise the value is 0. Each flow can only be transmitted in one of these ways, or give up transmission. Thus
	\begin{equation}
		a_{f}^{s}+a_{f}^{l}+a_{f}^{r}+a_{f}^{u}=\left\{ 0,1 \right\},\ \forall\; f.\label{CONS1}
	\end{equation}
	

There is no left-end relay mode for the leftmost MR and no right-end relay mode for the rightmost MR. So the following constraints exist,
	\begin{equation}
		\left\{
		\begin{array}{rcl}
			a_{f}^{l}=0,&& {{\rm{if}}\ f=1},\\
			a_{f}^{r}=0,&& {{\rm{if}}\ f=F}.
		\end{array}
		\right.\label{CONS2}
	\end{equation}
	
Define a blocked nodes set $\mathbb{B}$. When node $f$ receives the direct link power from the BS ,which is lower than the certain threshold $\varepsilon$, we identify it as a blocked node and put it into blocked set $\mathbb{B}$, which means that it can only use relays for transmission. Then
\begin{equation}
	f\in \mathbb{B},\;\;{\rm{if}}\ {{P}_{r}}\left( f \right)\le \varepsilon.\label{CONS3}
\end{equation}

If $f$ is a blocked node, it indicates that $f$ itself cannot select direct link for transmission, $a_{f}^{s}=0$. At the same time, the blocked node affects nearby MRs, so that it cannot be used to assist adjacent MRs. As a result, the node ${{MR}_{f-1}}$ at the left end cannot select its right-end MR for relaying, and the node ${{MR}_{f+1}}$ at the right end cannot select its left-end MR for relaying. Then
	\begin{equation}
		a_{f}^{s}=0\;\&\;a_{f-1}^{r}=0\;\&\;a_{f+1}^{l}=0,\;\;{\rm{if}}\ f\in \mathbb{B}.\label{CONS4}
	\end{equation}
	
After determining the transmission mode of the flow, the actual SINR and actual transmission rate of each flow can be obtained according to the values of binary variables $a_{f}^{s}$, $a_{f}^{l}$, $a_{f}^{r}$ and $a_{f}^{u}$. 
Then, $ SINR_{f} $ and $ {{R}_{f}} $ can be expressed as
	\begin{equation}
		\begin{aligned}
			SINR_{f}&=a_{f}^{s}\Gamma_{f}^{s}+a_{f}^{l}\Gamma_{f}^{l}+a_{f}^{r}\Gamma_{f}^{r}+a_{f}^{u}\Gamma_{f}^{u}.\\
			{{R}_{f}}&=a_{f}^{s}R_{f}^{s}+a_{f}^{l}R_{f}^{l}+a_{f}^{r}R_{f}^{r}+a_{f}^{u}R_{f}^{u}.
		\end{aligned}
		\label{CONS5}
	\end{equation}

Define a binary variable ${{C}_{f}}$ for each flow to indicate whether the flow $f$ is successfully scheduled within the specified time slots. If it is scheduled, the value is 1; otherwise, it is 0. Only when the QoS requirement and channel quality requirement are met, and the addition of flow $ f $ makes the total transmission time slots of all flows in the scheduling set not exceed the specified total time slots $ M $, this flow can be scheduled. Then the following constraints can be obtained,
	\begin{equation}
		{{C}_{f}}=1,\;\;
		{\rm{if}}\;
		\left\{
		\begin{aligned}
			SINR_{f}&\ge SIN{{R}_{f\min }},\\
			{{R}_{f}}&\ge {{r}_{f}},\\
			\sum\limits_{f=1}^{F}{{{C}_{f}}{{\delta }^{f}}}&\le M.
		\end{aligned}
		\right.\label{CONS6}
	\end{equation}

Using the constraints of (\ref{CONS1})--(\ref{CONS6}), the optimization problem (P1) of maximizing the number of flows transmitted within a finite time slots can be expressed as follows
	\begin{equation}\hspace{-2.8cm}
({\rm{P1}})\ \ \max \sum\limits_{f=1}^{F}{{{C}_{f}}}\\
	\end{equation}
	\hspace{2.8cm}s.t.
	\hspace{0.15cm}Constraints (\ref{CONS1})--(\ref{CONS6}).\\
This is a mixed integer nonlinear programming (MINLP) optimization problem. Because of its high computational complexity, this NP-hard problem urgently needs a solution with low complexity. Next, we plan to use heuristic algorithm to design an appropriate UAV relay assistance scheme and transmission scheduling scheme.

\section{Uav and MRs relay assistance algorithm}\label{S5}


In this section, we introduce a relay selection method for blocked links based on graph theory. The proposed heuristic UMRA algorithm is mainly to reduce the complexity of the original problem. The specific process is shown in Fig. \ref{fig:liucheng}. Firstly, the idea of blockage vector graph is introduced in \ref{S5-1} and applied to the following relay decision algorithm in \ref{S5-2}, which can make effective relay selection for blocked links and links that do not meet the requirements. Next, we describe the transmission scheduling algorithm in \ref{S5-3}, which completes the scheduling of as many transmission flows as possible within the specified number of time slots. Through a series of variable inputs, the final output is the number of flows and system throughput achieved by the UMRA algorithm. 

\subsection{Blockage Directed Vector Graph}\label{S5-1}
In order to solve the above optimization problem with constraints, we propose a method based on directed vector graph to generate the blockage vector graph of blocked nodes in advance. It reasonably predicts the remaining transmission modes of all links, reduces the impact of blocked nodes on other links to be transmitted, and quickly selects the appropriate relay mode for each flow. We will describe this idea through the following example.

\begin{figure}[t]
	\begin{minipage}[t]{1\linewidth}
		\centering
		\includegraphics[width=0.8\columnwidth]{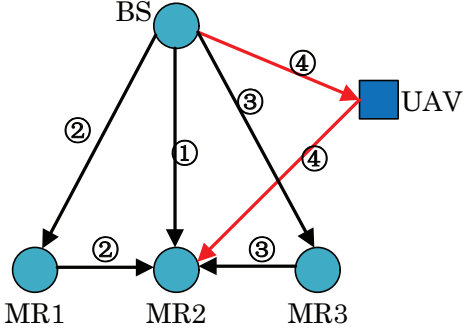}
	\end{minipage}%
	\caption{Transmission modes of any node}
	\label{fig:relay} 
	\vspace*{-3mm}
\end{figure} 

We consider that any MR can be transmitted in many ways. In Fig. \ref{fig:relay}, for $ \rm{MR_2} $, there are four transmission modes, via direct link transmission, i.e., $\rm{BS}\to M{{R}_{2}}$, left-end MR relay assistance transmission, i.e., $\rm{BS}\to \rm{MR_1}\to \rm{MR_2}$, right-end MR relay assistance transmission, i.e., $\rm{BS}\to \rm{MR_3}\to \rm{MR_2}$, UAV relay assistance transmission, i.e., $\rm{BS}\to \rm{UAV}\to \rm{MR_2}$. But for the leftmost $\rm{MR_1}$ and rightmost $\rm{MR_3}$, there are only three transmission modes.





\begin{figure}[t]
	\begin{minipage}[t]{1\linewidth}
		\centering
		\includegraphics[width=0.8\columnwidth]{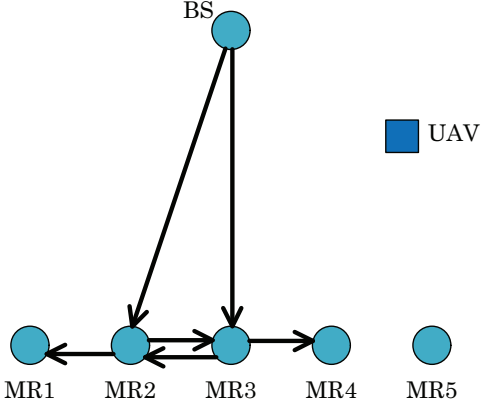}
	\end{minipage}%
	\caption{Directed blockage graph $G(V,E)$}
	\label{fig:zuse} 
	\vspace*{-3mm}
\end{figure}

Aiming at the problem of links blocked caused by obstacles in the actual transmission process, this paper proposes a method of directed vector graph to mark the blocked link. The node blocked makes the corresponding link unable to exploit the direct transmission mode and cannot be the relay for nearby MRs. Next, we take a directed vector graph $G(V,E)$ with some blocked links as an example, as shown in Fig. \ref{fig:zuse}.

In Fig. \ref{fig:zuse}, there are five MRs, one UAV and one BS. Assuming that the two direct links $ \rm{BS}\to \rm{MR_2} $ and $\rm{BS}\to \rm{MR_3}$ are blocked by obstacles. We put nodes $\rm{MR_2}$ and $\rm{MR_3}$ into the blocked nodes set $ \mathbb{B} $ , and establish directed segments from $ \rm{BS} $ to $\rm{MR_2}$ and $\rm{MR_3}$. Let
\begin{equation}
	\mathcal{L}_{BS,M{{R}_{2}}}^{e}=1\ \&\ \mathcal{L}_{BS,M{{R}_{3}}}^{e}=1.
\end{equation}

It shows that the $ \rm{BS} \to \rm{MR_2} $ and $ \rm{BS}\to \rm{MR_3}$ links in the graph $G(V,E)$ are blocked, and the blocked node cannot serve the nearby MR. Therefore, directed vectors are emitted from $\rm{MR_2}$ and $\rm{MR_3}$, informing nearby nodes that ``I can't do it''. Then, we set
\begin{equation}
	\begin{aligned}
		&{\mathcal{L}_{M{{R}_{2}},M{{R}_{1}}}^{e}=1\ \&\ \mathcal{L}_{M{{R}_{2}},M{{R}_{3}}}^{e}=1},\\
		&{\mathcal{L}_{M{{R}_{3}},M{{R}_{2}}}^{e}=1\ \&\ \mathcal{L}_{M{{R}_{3}},M{{R}_{4}}}^{e}=1}.\\
	\end{aligned}
\end{equation}

A new vector pointing to itself is generated at $\rm{MR_1}$, $\rm{MR_2}$, $\rm{MR_3}$ and $\rm{MR_4}$. After the directed graph is updated, there are only two ways for the blocked node $\rm{MR_2}$: its left-end MR relay and UAV-assisted relay. The blocked node $\rm{MR_3}$ has two modes: its right-end MR relay and UAV-assisted relay. $\rm{MR_1}$ can choose direct link or UAV-assisted relay, $\rm{MR_4}$ can select direct link, $\rm{MR_5}$ relay or UAV-assisted relay. $\rm{MR_5}$ is the rightmost node, there is no right-end MR relay. It has no node pointing to itself, therefore, direct link, left-end MR relay and UAV-assisted relay can be selected. When choosing the final transmission mode, it is more convenient for us to directly use the direct transmission link for $\rm{MR_1}$, $\rm{MR_4}$ and $\rm{MR_5}$ nodes that are not blocked.

\begin{algorithm}[t]
	\DontPrintSemicolon
	\caption{UAV \& MRs Relay Decision Algorithm}\label{alg:relay}
	\textbf{Input:} location information of BS, UAV and MRs, the minimum SINR requirement ${{\left( SINR \right)}_{f\min }}$ and QoS requirement ${{q}_{f}}$ of each flow $f$;\\
	\textbf{Initialization:} $G(V,E)=\emptyset$; $\mathbb{B}=\emptyset$; $\mathbb{S}, \mathbb{L}, \mathbb{R}, \mathbb{U}=\emptyset$; \\
	\For{$1\le f\le F$}
	{
		\If {${{P}_{r}}\left( f \right)\le \varepsilon $}
		{
			$\mathbb{B}=f\cup \mathbb{B}$;\\$\mathcal{L}_{BS,f}^{e}=1$; $\mathcal{L}_{f,f-1}^{e}=1$; $\mathcal{L}_{f,f+1}^{e}=1$;
		}
	}
    Form a directed vector graph $G(V,E)$ with blocked links .\\
	\For{$1\le f\le F$}
	{
		\If{$f\notin \mathbb{B}\;\;\&\;{{\left(SINR \right)}_{fs}}\ge {{\left( SINR \right)}_{f\min }}\;\&\;{{R}_{fs}}\ge {{r}_{f}}$}
		{The flow meeting the requirements can be transmitted through the direct link, and the node $ f $ is put into the set $\mathbb{S}$;\\
			${{R}_{f}}={{R}_{fs}}$;\\
			${{\left( SINR \right)}_{f}}= {{\left( SINR \right)}_{fs}}$;\\
			$\mathbb{S}=f\cup \mathbb{S}$;
		}
		\Else
		{
			The relay mode is adopted for transmission scheduling;\\
			\If{\emph{in graph} $G(V,E)$, \emph{there is a vector pointing to node} $ f $, \emph{that is} $\exists\; \mathcal{L}_{f-1,f}^{e}=1$ \emph{or} $\mathcal{L}_{f+1,f}^{e}=1$}
			{
				Abandon the corresponding relay mode, let ${{\left( SINR \right)}_{fl/r}}=0\;\&\; {{R}_{fl/r}}=0$;
			}
			Calculate the achievable rates and SINRs of all remaining relay modes of node $ f $ in graph $ G $;\\
			${{R}_{f}}=\max \left\{ {{R}_{fl}},{{R}_{fr}},{{R}_{fu}} \right\}$;\\
			$SIN{{R}_{f}}=\max \left\{ SIN{{R}_{fl}},SIN{{R}_{fr}},SIN{{R}_{fu}} \right\}$;\\
			\If{${R}_{f} \ge {r}_{f}\;\&\;{SINR}_{f} \ge {SINR}_{fmin}$}
			{
				Then choose the corresponding scheduling mode and put node $ f $ into the corresponding node set $ \mathbb{L} $, $ \mathbb{R} $ or $ \mathbb{U} $;
			}
			\Else {Abandon the scheduling of this flow;}
		}
	}
\end{algorithm}

\subsection{Relay Decision Algorithm}\label{S5-2}
The corresponding node of any blocked link is called a blocked node. Each blocked node has several relay modes, but not every relay mode can meet the QoS requirement and channel quality requirement. The blocked node cannot use the direct link to transmit itself, nor can it provide services for adjacent nodes. The flow that can be directly transmitted by the BS needs to meet the following three conditions at the same time:

1) The corresponding node is a non-blocked node.

2) Meet the QoS requirement.

3) Meet the requirement of channel quality (i.e., the minimum SINR value).

The relay mode shall be reasonably selected for other flows to realize the maximum rate of data transmission, so as to maximize the number of flows and throughput in the specified number of time slots. Aiming at how to adopt the appropriate link transmission mode, we propose a heuristic relay decision algorithm, and the idea of blockage vector graph is applied to it.

The blockage vector graph $G(V,E)$ contains the blockage relationship among all nodes. The edges in $G$ indicate unavailable relay modes. $ \mathbb{S} $, $ \mathbb{L} $, $ \mathbb{R} $ and $ \mathbb{U} $ are the link sets of the direct mode, left-end relay mode, right-end relay mode and UAV-assisted relay mode obtained after judgment by the UMRA relay decision algorithm. In this relay decision algorithm, if the blocked node cannot use MRs for relay, in order to reduce the number of flows which give up being scheduled, UAV-assisted relay and reasonable deployment location can greatly enhance the robustness of the train-ground communication system.

The heuristic relay decision algorithm using UAV and MRs is shown in the Algorithm \ref{alg:relay}. Lines 1--7 incorporate the relevant algorithm of blockage vector graph. In lines 1--2, we first input the location information of the BS, MRs and UAV, determine the minimum SINR requirement and QoS requirement of each flow, and initialize the blockage vector graph $G(V,E)$, the blocked nodes set $ \mathbb{B} $, the direct links set $ \mathbb{S} $, the left-end relay set $ \mathbb{L} $, the right-end relay set $ \mathbb{R} $ and the UAV relay assistance set $ \mathbb{U} $.
Lines 3--7 judge each flow in turn to generate a blockage vector graph $G(V,E)$. If the received power of the flow is less than the threshold, it is determined that the flow is blocked, and the corresponding node is put into the blocked nodes set $ \mathbb{B} $. We generate vectors from this point to its adjacent left-end and right-end MRs in the blockage graph $G(V,E)$.

After generating the blockage vector graph, we should make relay decision, as shown in lines 8--12. If the destination node of the flow is not an element in the blocked nodes set $ \mathbb{B} $, as well as SINR and transmission rate meet the requirements, the flow can be scheduled and put into the direct links set $ \mathbb{S} $. Otherwise, the relay decision will be utilized. As shown in lines 14--23, for each flow to be relayed, if there is a vector pointing to this MR in the blockage vector graph $ G(V,E) $, we should abandon the selection of this relay mode, and set the actual SINR and transmission rate of this relay mode to 0 to avoid affecting subsequent relay mode selections. Then calculate the achievable rate and SINR of all remaining relay modes of the node in the graph $ G(V,E) $. In lines 18--19, we choose the relay mode with the largest SINR and transmission rate. Judge whether this relay mode meets the minimum requirements of SINR and transmission rate through lines 20--23. If it meets the requirements, this relay mode will be exploited and the node will be placed in the corresponding node set; otherwise, the transmission of this flow will be abandoned.

For the heuristic relay decision algorithm, in Line 3, we traverse all MR nodes to determine whether they are blocked. The 
$ for $ loop has F iterations. Hence, the complexity of the relay decision algorithm is $\mathcal{O(\text{F})}$. In order to reduce the computational complexity, in the relay selection algorithm, we believe that each flow can only select one way to be transmitted and scheduled. Therefore, in Line 8, we traverse all nodes and let the flow with traffic choose a way to schedule. The $ for $ loop is the same as that in Line 3, and the two are in parallel relationship, so the complexity of algorithm 1 is $\mathcal{O(\text{F})}$.
	
\subsection{Scheduling Algorithm}\label{S5-3}
After reasonably selecting the transmission mode for each flow, a heuristic transmission scheduling algorithm is proposed. As shown in the constraint condition(\ref{CONS6}), only when adding this flow that satisfies the QoS requirement and the required transmission rate ${{r}_{f}}$, the total number of time slots currently spent does not exceed the specified number of transmission time slots $ M $, it can be successfully scheduled. In order to achieve a higher number of completed flows and system throughput, the scheduling priority needs to be set for each flow, and finally all flows are scheduled according to the priority order.
	
In this algorithm, the priority $ {\psi }_{f} $ is defined as the reciprocal of the number of time slots spent by flow $ f $, which can be expressed as
	\begin{equation}
		{{\psi }_{f}}=\frac{1}{{{\delta }^{f}}}=\left\lfloor \frac{{{R}_{f}}*\Delta t}{{{q}_{f}}\left( {{T}_{s}}+M*\Delta t \right)} \right\rfloor. 
	\end{equation}
The more time-consuming the flow is, the lower its priority and the less likely it is to be scheduled. Flows that spend a small number of slots are scheduled ahead to make better use of limited resources to maximize the scheduled throughput and the number of flows.

\begin{algorithm}[t]
	\DontPrintSemicolon
	\caption{Transmission Scheduling Algorithm}\label{alg:scheduling}
	\textbf{Input:} the set of $ \mathbb{S} $, $ \mathbb{L} $, $ \mathbb{R} $, $ \mathbb{U} $, the actual transmission rate ${{R}_{f}}$ of each flow $f$; \\
	\textbf{Output:} number of successfully scheduled flows A, state ${{C}_{f}}$ of each flow, total throughput I; \\
	\textbf{Initialization:} ${{C}_{f}}=0$; $\xi =0$; $A=0$; $I=0$; \\
	Calculate the number of time slots required for each link according to the set of $ \mathbb{S} $, $ \mathbb{L} $, $ \mathbb{R} $ and $ \mathbb{U} $ $     {{\delta }_{f}}=\left\lceil \frac{{{q}_{f}}\left( {{T}_{s}}+M*\Delta t \right)}{{{R}_{f}}*\Delta t} \right\rceil $.\\
	Arrange and schedule all flows in ascending time order.\\
	\For{${\rm{slot}}\;\; t (1\le t\le M)$}
	{
		\If{$\xi =0$}
		{
			Then a new flow $f$ is added in time slot order;\\
			Let ${{I}_{f}}={{q}_{f}}\left( {{T}_{s}}+M*\Delta t \right)$\ and \ $\xi =1$;
		}
		\If{$\xi =1$}
		{
			${{I}_{f}}={{I}_{f}}-{{R}_{f}}*\Delta t$;\\
			$I=I+{{R}_{f}}*\Delta t$;\\
			\If{${{I}_{f}}\le 0$}
			{
				$A=A+1$; ${C}_{f}=1$; $\xi =0$;
			}
		}
	}
\end{algorithm}	

The scheduling algorithm is shown in the Algorithm \ref{alg:scheduling}. In lines 1--2, we input four sets of $ \mathbb{S} $, $ \mathbb{L} $, $ \mathbb{R} $ and $ \mathbb{U} $ judged by the relay selection and the actual transmission rate ${{R}_{f}}$ of each flow. We finally output the number of successfully scheduled flows $ A $, the state ${{C}_{f}}$ of each flow and the total throughput $ I $.

In lines 4--5, we first integrate four sets $ \mathbb{S} $, $ \mathbb{L} $, $ \mathbb{R} $ and $ \mathbb{U} $ of flows to be transmitted. According to the QoS requirements and the actual transmission rates, we calculate the number of time slots $ {\delta}^{f} $ required for each link. In order to maximize the number of transmission flows in the specified time slots, these flows are scheduled in ascending order of the number of required time slots, so that flows with fewer time slots are scheduled first. $ M $ time slots are divided into several stages. Each stage contains several time slots and can only serve one flow. Then, the flows to be scheduled are scheduled in order. As shown in lines 6--14, within the specified number of time slots, the flows are scheduled in the order of priority $ {\psi }_{f} $. If the link residual traffic ${{I}_{f}}\le 0$, it indicates that the flow has been fully scheduled, and the flow state ${{C}_{f}}$ is set to 1. After any flow is scheduled, a new flow will be added to be scheduled.

In Line 6, there is a $ for $ loop with M iterations. Hence, the complexity of the scheduling algorithm is $\mathcal{O(\text{M})}$.
	\begin{table}[t]
	\caption{Simulation parameters} \label{table:parameter setting}
	\centering  
	\begin{tabular}{lccc}
		\hline
		\textbf{Parameter}  &\textbf{Symbol}&\textbf{Value}\\
		\hline
		Transmitting power &$ {P}_{t} $&1000mW\\
		Transmission frequency &$ f $&28GHz\\
		Bandwidth&$ W $&1200MHz\\
		Background noise&$ {N}_{0} $&-134dBm/MHz\\
		Path loss index&$ n $&2\\
		Half power beam width&$ \theta_{\mbox{\scriptsize{-3dB}}} $&$ 30^\circ $\\
		Slot time&$ \Delta t $&$ 18\mu s $\\
		Scheduling phase time&$ {T}_{s} $&$ 850\mu s $\\
		SI cancellation parameter& $\beta$ & ${10^{-13}}$\\
		The height of UAV&$ {h}_{UAV} $& 100m \\
		The height of BS&$ {h}_{BS} $& 10m \\
		The height of MR&$ {h}_{MR} $& 2.5m \\
		\hline
	\end{tabular}\\
\end{table}

\section{Performance Evalution}\label{S6}
In this section, we compare the proposed algorithm with the optimal solution and benchmark schemes, and evaluate the system performance of the UMRA algorithm under different conditions.

\subsection{Simulation Setup}\label{S6-1}
In the simulation, the train has eight carriages with a total length of 200 m. 16 MRs are evenly distributed on the rooftop of the train, and there are several flows to be transmitted. The QoS requirement of each flow is randomly generated between 10 Mbps and 40 Mbps. The UAV is located 40 meters from the horizontal position relative to the front of the train. We need to complete as many flows as possible within $ 2.4 \times {10^3} $ transmission time slots. For FD communication, the higher the SI cancellation level, the smaller the $ \beta $ parameter, and the easier it is for the flow to meet the transmission requirements. Here, we set the SI cancellation parameter $ \beta $ to -130 dB. Other parameters are shown in Table \ref{table:parameter setting}.

The system adopts the real directional antenna model in IEEE 802.15.3c standard, with a Gaussian-shaped main lobe and constant-gain side lobe. The antenna gain $ G(\theta $) is expressed as 
	\begin{equation}
		G(\theta) =
		\begin{cases}
			G_0-3.01\times\left(\frac{2\theta}{\theta_{\mbox{\tiny{-3dB}}}}\right)^2, &\mbox{$0^{\circ}\le\theta\le\theta_{ml}/2$},\\
			G_{sl}, &\mbox{$\theta_{ml}/2<\theta\le180^{\circ}$},
		\end{cases}
	\end{equation}
where $ {\theta} $ takes a value in  $[0^\circ,180^\circ]$. $ {\theta_{\mbox{\tiny{-3dB}}}} $ is the half-power beamwidth. The main lobe width $ \theta_{ml} $ can be expressed as $\theta_{ml}=2.6\times\theta_{\mbox{\scriptsize{-3dB}}}$. $G_0$ is the maximum antenna gain, which can be expressed as
\begin{equation}
	G_0=10\mbox{log}(1.6162/\mbox{sin}(\theta_{{\mbox{\scriptsize{-3dB}}}}/2))^2.
\end{equation}
 $ G_{sl} $ is the side lobe gain, which can be expressed as 
 \begin{equation}
 	G_{sl}=-0.4111\times\mbox{ln}(\theta_{\mbox{\scriptsize{-3dB}}})-10.579.
 \end{equation}

To demonstrate the improvement in the performance of the heuristic UMRA algorithm proposed in this paper, in terms of the number of completed flows and throughput, we compare the heuristic algorithm with the other two benchmark schemes:
	
1) \emph{\textbf{MRA}}: There is no UAV-assisted for relay. The optimal relay mode is selected using a relay-assisted algorithm of MRs deployed on top of the train, and then we carry out transmission and scheduling.
	
2) \emph{\textbf{RA}}: The hybrid relay of MRs deployed on the train and UAV adopts a unified way, arbitrarily selects the relay mode and then we carry out the transmission scheduling algorithm. 
	
\subsection{Compared With the Optimal Solution}\label{S6-2}
\begin{figure}[t]
		\begin{minipage}[t]{1\linewidth}
			\centering
			\includegraphics[width=2.9in]{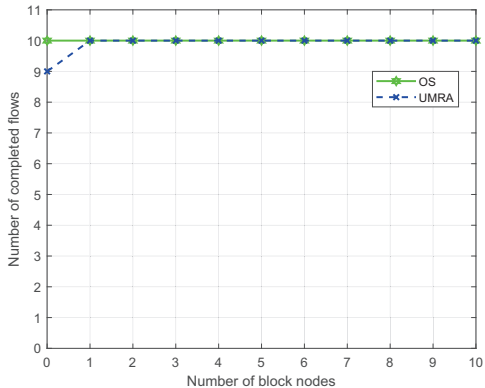}
		\end{minipage}%
		\caption{Comparison with the exhaustive search algorithm}
		\label{fig:os 1400} 
		\vspace*{-3mm}
	\end{figure}

In this subsection, we compare the results of the heuristic UMRA algorithm with the optimal solution (OS) obtained by the exhaustive search algorithm. Due to the high computational complexity of the exhaustive search method, we reduce the number of flows to be transmitted for simulation. It is assumed that there are 10 data flows to be transmitted, throughputs are randomly selected between 10 Mbps and 40 Mbps, and all flows are planned to complete transmission within $ 1.4 \times {10^3} $ time slots. We place the UAV 40 meters away from the horizontal position of the railway engine. To test the heuristic algorithm in all link environments. We simulate the number of blocked nodes from 0 to 10. In the worst case, all nodes are blocked. Fig. \ref{fig:os 1400} shows the comparison of the number of flows completed by the UMRA algorithm and the exhaustive algorithm.
	
	The average deviation can be calculated as 
	\begin{equation}
		Average\      Deviation=\frac{1}{10}\sum\limits_{n=1}^{10}{\frac{{{C}_{OS}}\left( n \right)-{{C}_{UMRA}}\left( n \right)}{{{C}_{OS}}\left( n \right)}}.
	\end{equation}	
$ {{{C}_{OS}}\left( n \right)} $ and $ {{{C}_{UMRA}}\left( n \right)} $ represent the number of flows completed by the exhaustive algorithm and heuristic UMRA algorithm in the case of different numbers of blocked nodes, respectively. As we can see, using the same number of time slots, the number of flows completed by the UMRA algorithm approximates OS obtained by exhaustive algorithm. According to the above formula, the calculated average deviation is 1\%. Under the condition of reasonable SINR threshold, even if the link environment is bad, the UMRA algorithm still realizes the transmission of more data flows as much as possible and approaches the optimal solution. We can suggest that it is a suboptimal algorithm. 

For the exhaustive algorithm, we assume that there are $ f $ flows to be transmitted. Firstly, each flow with traffic to be transmitted needs to calculate the transmission rates of four transmission modes, this process requires F iterations and the complexity is $\mathcal{O(\text{F})}$. In the exhaustive scheduling algorithm, each flow has five options. Each flow with traffic needs to traverse these five options. The outer loop is $ f $ nested for loops, and the inner loop is 5, so the complexity is $\mathcal{O(\text{$ 5^f $})}$. The worst case is that each MR has a flow to be transmitted, the complexity of the exhaustive algorithm is $\mathcal{O(\text{$ 5^F $})}$. In the transmission process, the traffic transmission needs to be completed within the specified number of time slots. So the $ for $ loop requires M iterations, and the complexity is $\mathcal{O(\text{M})}$. It can be seen that with the increase of the number of flows to be transmitted, the advantages of the UMRA algorithm are more obvious. Therefore, the heuristic algorithm realizes that when the performance of the two is highly similar, the computational complexity is greatly reduced, and ensures the efficient and robust transmission of the train-ground communication system.
	
\subsection{Simulation Results}\label{S6-3}
\begin{figure}[t] 
	\centering
	\subfigure[Number of completed flows]{
		\includegraphics[width=2.9in]{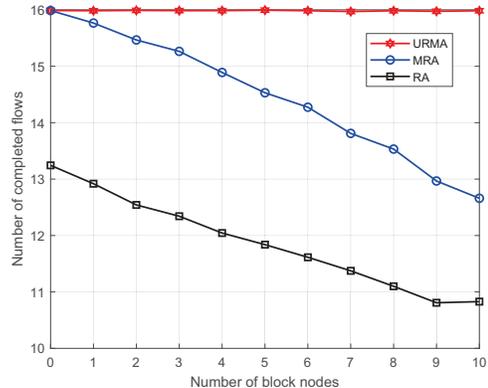}}
	\centering
	\subfigure[System throughput]{
		\includegraphics[width=2.9in]{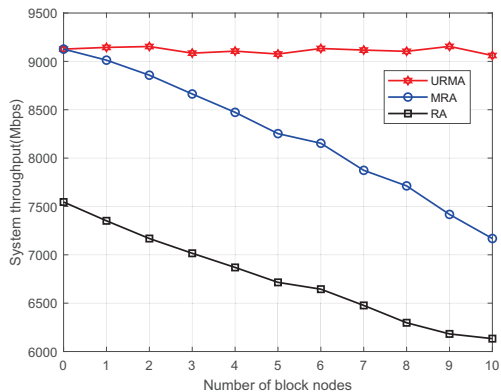}}
	\caption{Influence of the number of blocked nodes on system performance (with 16 transmission flows)}
	\label{fig:block}
\end{figure}

Firstly, the impact of the number of blocked nodes on the system performance is analyzed. It is assumed that there are 16 data flows to be transmitted, and the number of transmission time slots is $ 2.4 \times {10^3} $. The minimum SINR is set to $ 7 \times {10^4} $, as shown in Fig. \ref{fig:block}.
	
With the increase of the number of blocked nodes, the advantages of the UMRA algorithm are more obvious, and UMRA can complete more flows to be transmitted. In terms of the number of flows, UMRA can transmit an average of 15.98 flows, MRA can transmit an average of 14.47 flows, and RA can transmit an average of 11.88 flows. Therefore, the performance of UMRA is 10.44\% higher than that of MRA and 34.51\% higher than that of RA. In terms of throughput, the average throughputs of the UMRA, MRA and RA algorithm are $ 9.115 \times {10^3} $ Mbps, $ 8.246 \times {10^3} $ Mbps and $ 6.764 \times {10^3} $ Mbps. Therefore, the performance of the UMRA algorithm is 10.53\% higher than the MRA algorithm and 34.76\% higher than the RA algorithm.

We can see that when the blocked nodes are few and scattered, the role of UAV assistance is not obvious. 
This is because fewer blocked nodes will not have too much effect on the relay mode of nearby nodes. When more than half of the nodes are blocked, many nodes lose their way of relaying with nearby MR assistance. The MRA algorithm uses conventional nodes to relay, so that flows with severe congestion still do not meet the QoS requirements and cannot be placed in the pre-scheduled set. There are more remaining time slots, resulting in a waste of resources. Using a new type of air relay, obstacles can be bypassed. Appropriate UAV location enables more flows to meet QoS requirements and channel quality requirements. It is a good way to overcome large-area link blockage. Therefore, when large-area blockage occurs, UAV-assisted relay plays a great role, raising performance by 23.08\%.

	
	\begin{figure}[t] 
	\centering
	\subfigure[Number of completed flows]{
		\includegraphics[width=2.9in]{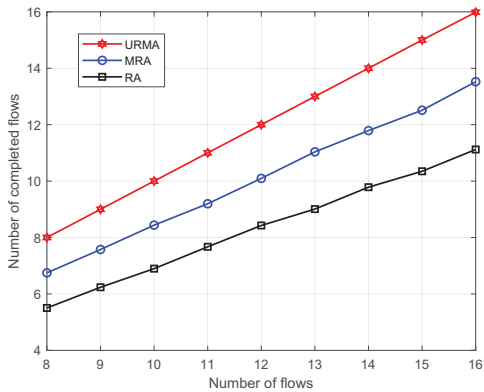}}
	
	\centering
	\subfigure[System throughput]{
		\includegraphics[width=2.9in]{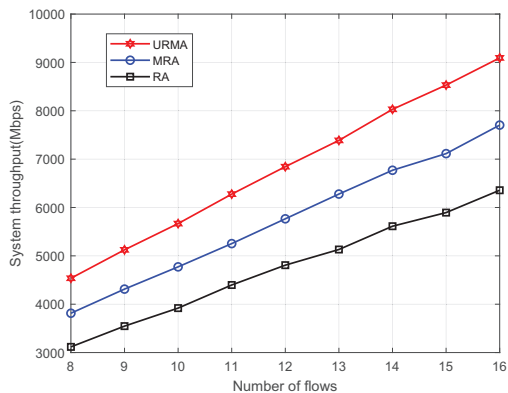}}
	\caption{Influence of the number of flows to be transmitted on system performance (with 8 blocked nodes)}
	\label{fig:flow}
\end{figure}	
	
In Fig. \ref{fig:flow}, we analyze the impact of the number of transmission flows on the system performance. In the simulation, we set the number of transmission slots to $ 2.4 \times {10^3} $ and the number of blocked nodes to 8, which means that half of the nodes are blocked. Generally speaking, with the increase of the number of flows to be transmitted, the number of successfully scheduled flows also increases. However, UMRA can complete more flows. For the number of flows, the performance of UMRA raises by 18.81\% compared with MRA and 44.06\% compared with RA. In terms of throughput, the performance of UMRA increases by 18.75\% compared with MRA and 43.72\% compared with RA. According to the trend of the curve, we can see that the more flows to be transmitted and the more blocked nodes, the more obvious the advantages of the UMRA algorithm.
	
\begin{figure}[t] 
	\centering
	\subfigure[Number of completed flows]{
		\includegraphics[width=2.9in]{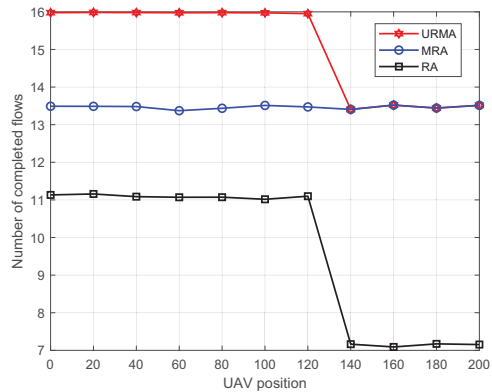}}
	
	\centering
	\subfigure[System throughput]{
		\includegraphics[width=2.9in]{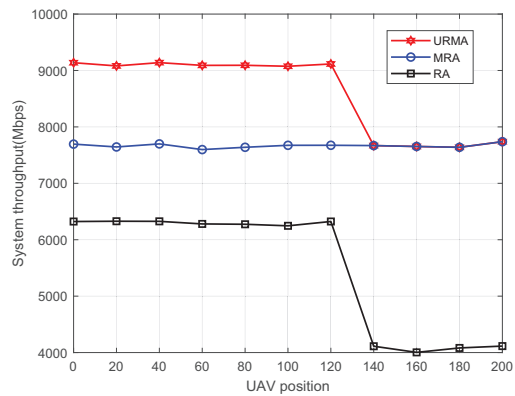}}
	\caption{Influence of UAV position on system performance (with 8 blocked nodes)}
	\label{fig:UAV}
\end{figure}

\begin{figure}[t]
	\begin{minipage}[t]{1\linewidth}
		\centering
		\includegraphics[width=2.9in]{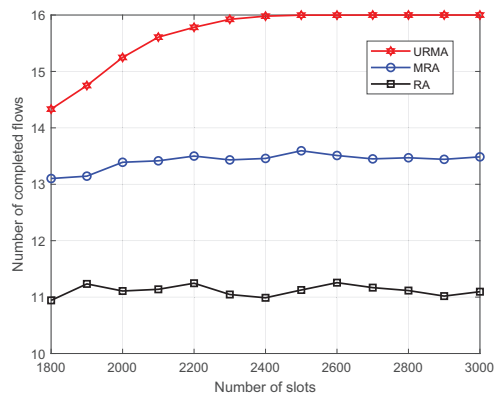}
	\end{minipage}%
	\caption{Influence of total transmission time slots on system performance (with 8 blocked nodes)}
	\label{fig:slot} 
	\vspace*{-3mm}
\end{figure}

Next, we simulate and analyze the deployment position of the UAV, as shown in the Fig. \ref{fig:UAV}. It is set as $ 2.4 \times {10^3} $ transmission time slots and the research scene has 8 blocked nodes. It is planned to complete the transmission of 16 flows. We find that the deployment position of UAV has a great impact on the performance. The farther the UAV is from the horizontal position of the locomotive, the less ideal the relay effect is. When it is more than 120 meters away from the locomotive, the performance decreases rapidly, and finally tends to MRA. However, no matter where the UAV is, the performance of the RA algorithm is much lower than that of UMRA and MRA. As can be seen from Fig. \ref{fig:UAV}(a), when the UAV position is set between 0-120 m in the horizontal position, the performance is the best. Combined with the throughput, in the simulation result Fig. \ref{fig:UAV}(b), when the horizontal position is set to 40 m, the number of completed flows and the system throughput are maximum.

\begin{figure}[t] 
	\centering
	\subfigure[Number of completed flows]{
		\includegraphics[width=2.9in]{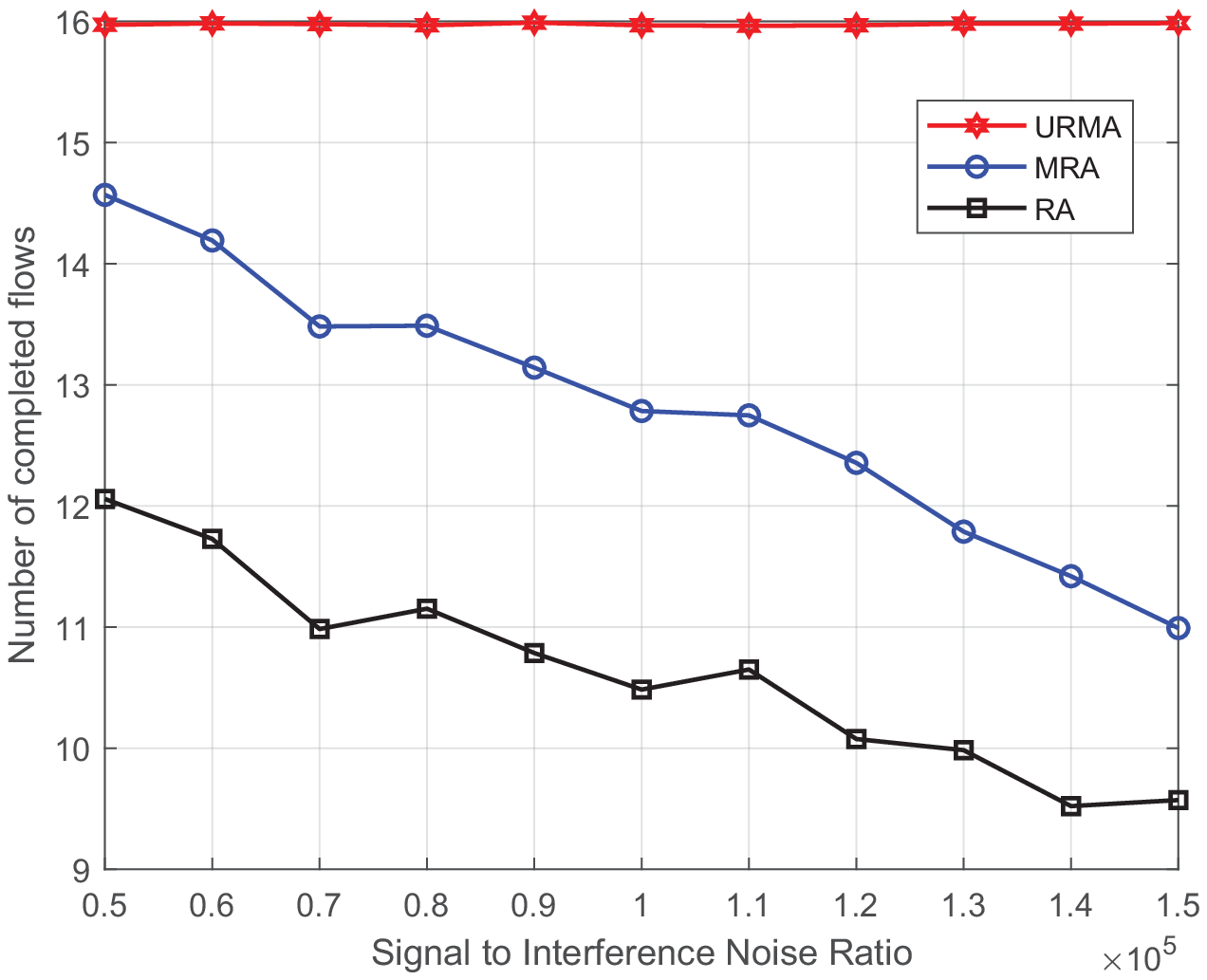}}
	
	\centering
	\subfigure[System throughput]{
		\includegraphics[width=2.9in]{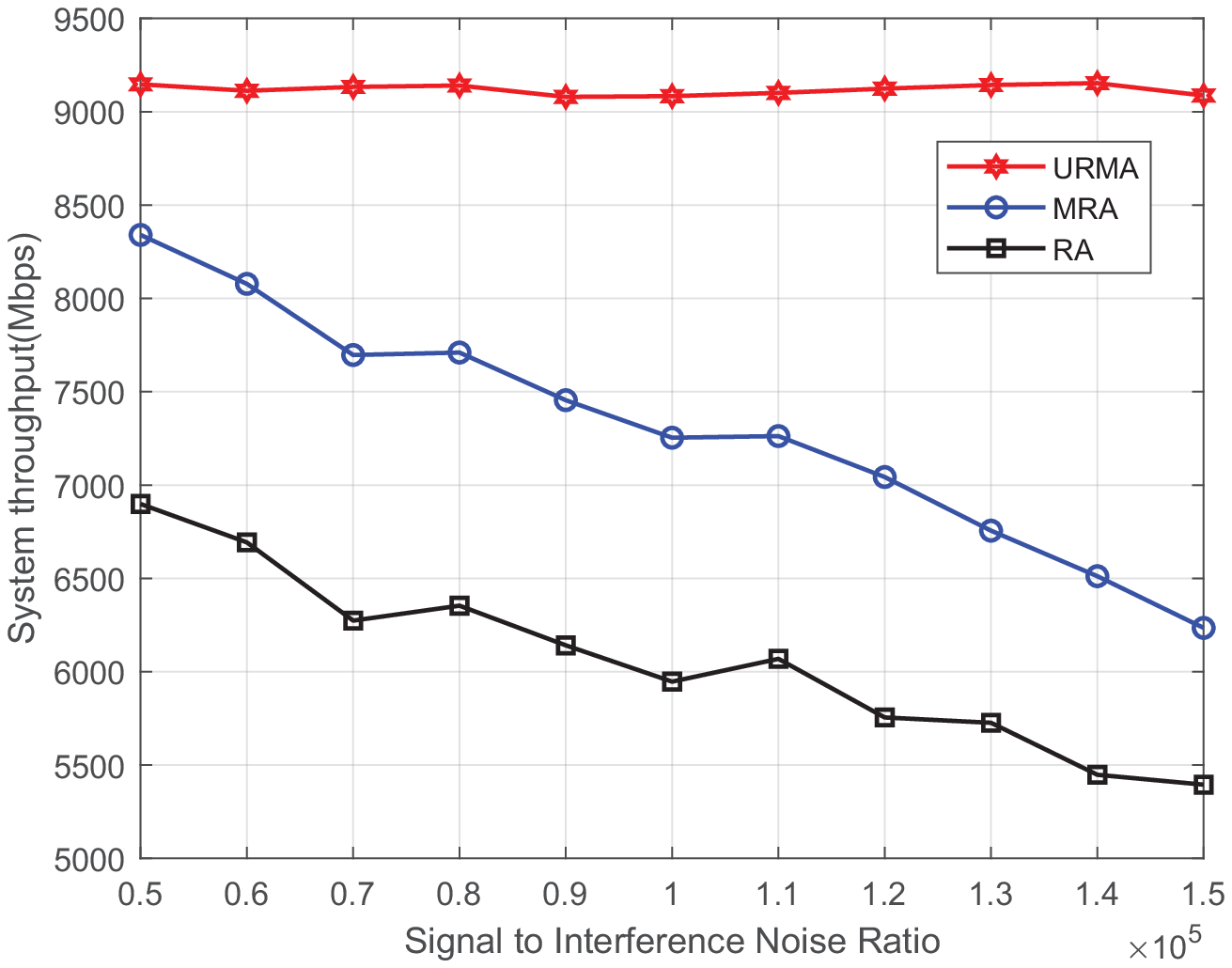}}
	\caption{ Influence of minimum signal-to-interference-plus noise ratio on
		system performance (with 8 blocked nodes)}
	\label{fig:SINR}
\end{figure}

In Fig. \ref{fig:slot}, it can be seen that when the total number of transmission time slots is $ 2.4 \times {10^3} $, the transmission of all flows is realized for the first time. With the increase of the number of time slots, the number of completed flows tends to be stable. It shows that appropriate transmission time slots can realize the full utilization of time resources, and achieve the best performance.
	
Assuming that there are $ 2.4 \times {10^3} $ transmission time slots. In Fig. \ref{fig:SINR}, when the minimum SINR requirement is between 0 and $ 9 \times {10^4} $, the transmission of all flows can be completed using the UMRA algorithm. With the improvement of the channel quality requirement, it is necessary to improve the minimum SINR. Generally, the number of flows meeting the requirements is less and less, which is fully reflected in the MRA and RA algorithms. However, the UMRA algorithm is hardly affected by SINR. This is because the use of UAV can introduce a new type of air link into the system, which can create a whole new channel environment to improve the situation of blockage by obstacles. By improving the aerial position of the UAV, we achieve a signal-to-noise ratio that cannot be achieved by traditional means. This will help flows that give up scheduling due to the link environment to a large extent. When the minimum SINR required by the channel exceeds $ 1.3 \times {10^5} $, the performance gaps between these three algorithms gradually increase, and the number of completed flows by UMRA exceeds 40\% of MRA and 65\% of RA.

\section{Conclusion}\label{S7} 

This paper proposes UMRA which is a heuristic robust algorithm using UAV and MRs relay assistance based on graph theory. It can reasonably select the relay mode of each link, effectively solve the obstacle blockage problem, fully schedule the flows that meet the QoS requirements and channel qualities, and greatly improve the number and throughput of completed flows. The simulation results show that the performance of UMRA approximates the optimal solution obtained by the exhaustive method, and the average deviation is 1.43\%. Compared with the other two benchmark schemes, when large-area blockage occurs, the number of flows completed by the UMRA algorithm raises by 23\%. With the increasing requirement for channel quality, the number of flows that UMRA can complete is 40\% higher than that of MRA. Therefore, the UMRA algorithm effectively improves the system performance while reducing the computational complexity.

\end{document}